\documentclass[a4paper,11pt]{article}
\usepackage{tikz}
\usetikzlibrary{arrows.meta}
\usepackage{amsmath}

\usepackage{aaskaiid}
\usepackage{aas_macros}%
\usepackage[utf8]{inputenc}
\usepackage{xspace}
\hypersetup{
    colorlinks=true,
    linkcolor=red,
    filecolor=magenta,      
    urlcolor=magenta,
    citecolor=blue,
}
\usepackage{multirow}
\usepackage{tabularx} 
\usepackage{adjustbox} 
\usepackage{pdflscape} 
\usepackage{graphicx} 
\usepackage{lscape} 
\usepackage{longtable} 
\usepackage{booktabs} 
\usepackage{caption} 
\usepackage{adjustbox}
\usepackage{threeparttable}
\usepackage{orcidlink}

\usepackage[utf8]{inputenc}
\usepackage{tikz}
\usetikzlibrary{arrows.meta,positioning,shapes.geometric}

\newcommand{\skalow}{SKA1-Low\xspace}
\newcommand{\skamid}{SKA1-Mid\xspace}
\newcommand{\rsun}{R_\odot}

\DeclareUnicodeCharacter{2033}{\textquotedbl}
\title{Solar, Heliospheric and Ionospheric Physics: Pathfinders, Precursors and SKAO Perspectives}
\ShortTitle{SHI science with SKAO}

\author[1]{Pietro Zucca\orcidlink{0000-0002-6760-797X}}
\ShortName{P. Zucca et al.} 
\author[2]{Rohit Sharma}
\author[3]{Divya Oberoi\orcidlink{0000-0002-4768-9058}}
\author[4]{Eduard Kontar\orcidlink{0000-0002-8078-0902}}
\author[5]{John Morgan\orcidlink{0000-0001-9224-5483}}
\author[6]{Peter Gallagher}
\author[7,8]{Sven Wedemeyer}
\author[9]{Biagio Forte}

\affiliation[1]{ASTRON - Netherlands Institute for Radio Astronomy, Oude Hoogeveensedijk 4, 7991 PD Dwingeloo, The Netherlands}
\emailAdd{zucca@astron.nl}

\affiliation[2]{Indian Institute of Technology, Kanpur 208016, India.}
\emailAdd{rsharma@iitk.ac.in}

\affiliation[3]{National Centre for Radio Astrophysics, Tata Institute of Fundamental Research, S. P. Pune University Campus, Pune, India}
\emailAdd{div@ncra.tifr.res.in, div.oberoi@gmail.com}

\affiliation[4]{Uni Glasgow}
\emailAdd{Eduard.Kontar@glasgow.ac.uk}

\affiliation[5]{CSIRO, Space and Astronomy, P.O. Box 1130, Bentley, WA 6102, Australia}
\emailAdd{john.morgan@csiro.au}

\affiliation[6]{Astronomy \& Astrophysics Section, School of Cosmic Physics, Dublin Institute for Advanced Studies, DIAS Dunsink Observatory, Dublin D15 XR2R, Ireland}
\emailAdd{peter.gallagher@dias.ie}

\affiliation[7]{Rosseland Centre for Solar Physics, University of Oslo, Oslo, Oslo, N-0315, Norway}
\affiliation[8]{Institute of Theoretical Astrophysics, University of Oslo, Oslo, Oslo, N-0315, Norway}
\emailAdd{sven.wedemeyer@astro.uio.no}

\affiliation[9]{Department of Electronic and Electrical Engineering
University of Bath
BA2 7AY, Bath, United Kingdom}
\emailAdd{b.forte@bath.ac.uk}

\abstract{The Solar, Heliospheric and Ionospheric (SHI) Physics Science Working Group of the
Square Kilometre Array Observatory (SKAO) addresses the full chain of plasma processes
linking the solar corona to the terrestrial environment. This overview chapter synthesises
16 topical contributions to \textit{Advancing Astrophysics with the SKA~-- II}, spanning
the quiet and active solar atmosphere, eruptive phenomena, heliospheric turbulence and
solar-wind diagnostics, ionospheric science, stellar--solar connections, and the
observational frameworks required to deliver these science goals. The primary focus is on
the capabilities of Array Assembly~4 (AA4), the design baseline for both \skalow{}
(50--350\,MHz) and \skamid (0.35--15.4\,GHz), which together provide continuous spectral
coverage, sub-arcsecond angular resolution, full-Stokes polarimetry, and sensitivity gains
of an order of magnitude over existing facilities. From resolving fine-scale coronal heating
events to mapping coronal mass ejection magnetic fields and characterising multi-scale
heliospheric turbulence, SKAO will deliver transformative advances in solar and space-weather
science. We frame these contributions as a single end-to-end Sun-to-Earth system. We
identify cross-cutting themes and gaps not fully addressed by individual chapters and outline
the staged roadmap from early operations through to the full AA4 capability.}

\begin{document}
\maketitle

\section{Introduction and Scope}
\label{sec:intro}

The Sun is the most powerful particle accelerator, the brightest radio source, and the
dominant driver of space weather in our local cosmic environment. Understanding the
physical processes that heat the million-degree corona, accelerate particles to
relativistic energies, launch coronal mass ejections (CMEs), and shape the heliospheric
plasma all the way to Earth's ionosphere remains among the central challenges in
astrophysics. Radio observations occupy a privileged position in this endeavour: the
emission mechanisms---thermal bremsstrahlung, gyroresonance and gyrosynchrotron radiation,
and coherent plasma emission---provide diagnostics of coronal magnetic fields, electron
energy distributions, and plasma turbulence that are inaccessible at other wavelengths.

The SKAO will deliver a step change in radio observing capability for SHI science.
\skalow, with 512 stations of 256 log-periodic dipole antennas spanning baselines up to
$\sim$65\,km, covers 50--350\,MHz and probes the corona and inner heliosphere from
roughly 1.1 to $>$3\,$\rsun$. \skamid, comprising 197 dishes (133 SKA dishes plus 64
MeerKAT dishes) with baselines exceeding 150\,km, covers 0.35--15.4\,GHz in four receiver
bands (Bands~1, 2, 5a, 5b), accessing the upper chromosphere, transition region, and low corona
with sub-arcsecond resolution. Together, the two instruments provide continuous frequency
coverage from 50\,MHz to 15.4\,GHz with full-Stokes spectropolarimetric imaging at
cadences from milliseconds to hours.

This volume presents the science case against the design baseline Array Assembly~4 (AA4),
with supplementary discussion of earlier array assemblies (AA$^*$, AA2) and future
enhancements. AA4 delivers the full collecting area, the densest instantaneous
$uv$-coverage, and the design sensitivity---$A_\mathrm{eff}/T_\mathrm{sys} \sim
1000$\,m$^2$\,K$^{-1}$ for \skalow{} at 110\,MHz and $\sim$1600\,m$^2$\,K$^{-1}$ for
\skamid at 1.4\,GHz---that underpin the science goals described in the following chapters.

The SHI contribution to \textit{Advancing Astrophysics with the SKA~-- II} comprises
16~topical chapters that may be grouped as follows: the solar atmosphere, from the quiet
corona through eruptive events \citep{Mondal01.2026.SKA, Dey01.2026.SKA, Kontar01.2026.SKA,
Sharma01.2026.SKA, Nakariakov01.2026.SKA, Patra01.2026.SKA, Kumari01.2026.SKA, Cheung01.2026.SKA,
Kansabanik01.2026.SKA}; heliospheric turbulence, solar wind, and energetic-particle
transport \citep{Chrysaphi01.2026.SKA, PeijinZhang01.2026.SKA, 
Morosan01.2026.SKA}; the ionosphere and geospace environment \citep{Datta01.2026.SKA,
Morgan01.2026.SKA}; stellar connections \citep{Mohan01.2026.SKA}; and the underpinning
observational, calibration, and data-analysis framework \citep{Oberoi01.2026.SKA}. This
overview chapter serves as the editorial thread that links these contributions, provides a
concise synthesis, identifies relevant gaps, and contextualises the collective effort within
broader trends in heliophysics.

\begin{figure}[t]
\centering
\includegraphics[width=\columnwidth]{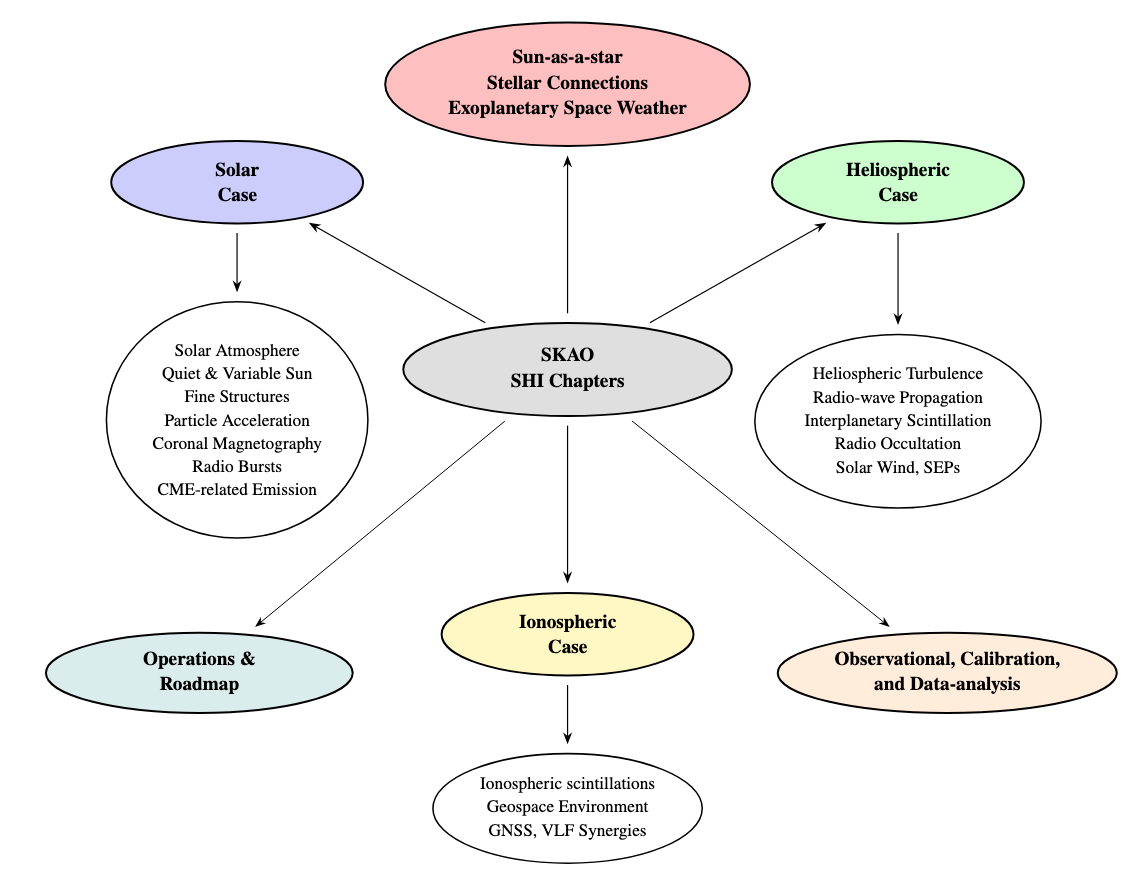}
\label{fig:diagram}
\caption{Overview of the 16 SHI topical chapters in \textit{Advancing Astrophysics with
the SKA~-- II} and their interconnections. Arrows indicate scientific dependencies. The observational framework
chapter of \citet{Oberoi01.2026.SKA} underpins all SHI science.}
\label{fig:overview_flowchart}
\end{figure}

These topics form a single end-to-end Sun-to-Earth system, in which energy and
disturbances propagate from coronal energy release (Section~\ref{sec:solar}), through the
turbulent heliosphere (Section~\ref{sec:heliosphere}), to their terrestrial imprint in the
ionosphere (Section~\ref{sec:ionosphere}); the scientific dependencies are summarised in
Figure~\ref{fig:overview_flowchart}.

\begin{figure}[t]
\centering
\includegraphics[width=\columnwidth]{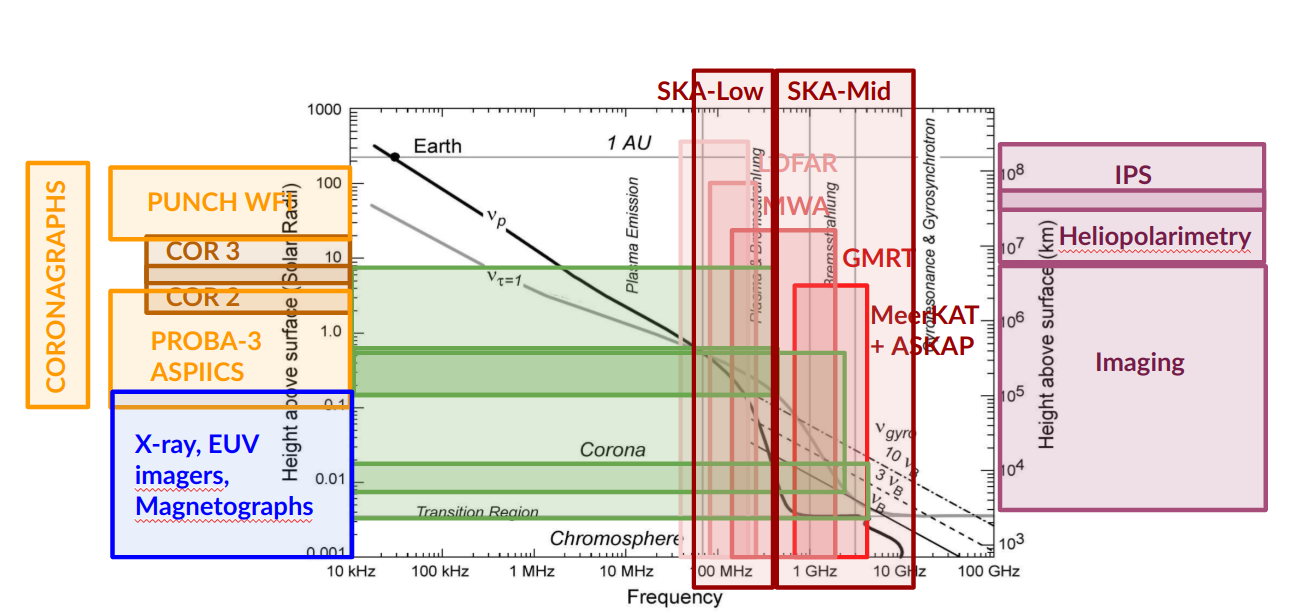}
\caption{Frequency--height diagram for solar radio observations with \skalow{} and \skamid.
The shaded regions indicate the approximate coronal and heliospheric heights probed at
each frequency via plasma emission (left axis), while the horizontal bars mark the
receiver bands and the angular resolution achieved at AA4 baselines. Emission mechanisms
transition from gyroresonance and bremsstrahlung at GHz frequencies to coherent plasma
emission below $\sim$300\,MHz. Adapted from Gary \& Hurford (2004).}
\label{fig:freq_height}
\end{figure}

\section{Solar Atmosphere -- From the Quiet Corona to Eruptive Events}
\label{sec:solar}

\subsection{Quiet Sun and Coronal Heating}
\label{sec:quietsun}

One of the oldest unsolved problems in solar physics is the mechanism by which the corona
is heated to temperatures exceeding $10^6$\,K. The nanoflare hypothesis, originally
proposed by \citet{Parker1988}, posits that ubiquitous small-scale impulsive energy
releases collectively supply the requisite energy budget. Recent observations with the
Murchison Widefield Array (MWA) have provided the first radio evidence for pervasive weak,
impulsive nonthermal emission from the quiet solar corona \citep{Mondal2020,2022ApJ...937...99S}, consistent
with this picture. \citet{Mondal01.2026.SKA} describe how SKAO's dramatic improvement in
snapshot dynamic range (exceeding $10^5$) and angular resolution at low frequencies will
enable statistical studies of these faint transients across the full solar disk. By
measuring the spectral and polarimetric properties of large populations of such events,
SKAO will quantify their nonthermal energy content and directly test whether small-scale
reconnection events can account for coronal heating.
The order-of-magnitude sensitivity gain lowers the weak-event detection threshold by a
comparable factor; given the steep event-energy power law, this yields population-scale
samples that can measure the distribution slope---the decisive test of the nanoflare
hypothesis.

\subsection{Coronal Seeing and Sub-arcsecond Dynamics}
\label{sec:seeing}

Radio images of the Sun are broadened by scattering in the turbulent corona, an effect
analogous to atmospheric seeing in optical astronomy. \citet{Dey01.2026.SKA} address this
``coronal seeing'' and its implications for high-resolution imaging with SKAO.
Observations with LOFAR and the MWA have revealed that some sources---notably type-I
noise-storm emission---are remarkably compact, with sizes as small as $\sim$9\,arcsec,
far smaller than previously thought possible at metre wavelengths. Understanding the
scattering kernel as a function of frequency, heliocentric distance, and heliographic
position is essential for deconvolving intrinsic source structures from
propagation-induced broadening. SKAO's long baselines and dense $uv$-coverage will
simultaneously constrain scattering models and reveal fine structures in the corona that
current instruments cannot resolve.

\subsection{Solar Radio Burst Fine Structures}
\label{sec:finestructures}

Solar radio bursts display a rich taxonomy of fine structures---narrowband spikes, stria
bursts, zebra patterns, fibre bursts---that encode information about the microphysics of
particle acceleration and the properties of the ambient plasma. \citet{Kontar01.2026.SKA}
review the diagnostic potential of these features and the requirements for exploiting them
with SKAO. Imaging spectroscopy with LOFAR has demonstrated that individual type-IIIb
stria burst elements can be spatially and spectrally resolved \citep{Kontar2017}, but the
limited $uv$-coverage restricts source characterisation. With \skalow, the combination of
sub-second time resolution, $\sim$10\,kHz spectral resolution, and imaging at arcminute to
sub-arcminute scales will enable routine identification and physical modelling of fine
structures across the full range of coherent emission processes.

\subsection{Particle Acceleration in Solar Flares}
\label{sec:particles}

Solar flares accelerate electrons and ions to high energies within seconds. The radio
signatures of these nonthermal populations span the full SKAO frequency range: coherent
plasma emission at low frequencies traces electron beams propagating through the corona,
while incoherent gyrosynchrotron emission at GHz frequencies maps the energy distribution
and pitch-angle anisotropy of trapped electrons in flaring loops.
\citet{Sharma01.2026.SKA} demonstrate that \skamid Band~5 observations (4.6--15.4\,GHz)
will resolve the spatial structure of gyrosynchrotron sources in individual flaring loops
with $\sim$0.1\,arcsec resolution, enabling spectral inversion to recover the magnetic
field strength, nonthermal electron density, and energy spectral index as functions of
position. When combined with \skalow{} observations of the associated coherent emission and
with hard X-ray data from missions such as Solar Orbiter STIX, SKAO will provide an
end-to-end characterisation of the acceleration and transport of energetic electrons in
flares.

\subsection{Quasi-Periodic Pulsations}
\label{sec:qpp}

Quasi-periodic pulsations (QPPs) are a common feature of solar flare emission, observed
across the electromagnetic spectrum from radio to gamma-rays. Their origin is debated:
competing models invoke magnetohydrodynamic (MHD) oscillations of flaring structures,
periodic or bursty reconnection, and nonlinear wave--particle interactions.
\citet{Nakariakov01.2026.SKA} describe how SKAO will advance QPP research by observing
simultaneously in both coherent and incoherent emission domains with continuous spectral
coverage. The high time resolution ($\lesssim$50\,ms) and sensitivity of SKAO will reveal
QPP signatures in weak flares that are presently undetectable, while machine-learning
methods applied to the high-cadence data streams will enable automated detection and
classification of QPP events across the full activity cycle.

\subsection{Coronal Magnetography}
\label{sec:magnetography}

Direct measurement of the coronal magnetic field---the fundamental quantity governing
coronal structure and dynamics---remains one of the principal observational challenges in
solar physics. Radio observations offer two complementary pathways: gyroresonance
emission, sensitive to the magnetic field strength in active-region loops at GHz
frequencies, and circular polarisation of plasma emission at low frequencies, encoding
the field in the emission and propagation regions. \citet{Patra01.2026.SKA} present
the case for routine coronal magnetography using SKAO's spectropolarimetric capabilities.
The exceptional polarisation purity of \skalow{} and \skamid{} ($\lesssim$0.1\% instrumental
polarisation leakage after calibration) will enable Stokes-$V$ measurements of thermal
gyroresonance layers in active regions with \skamid and broadband circular-polarisation
mapping of the quiet corona with \skalow, together yielding magnetic-field maps from the
low corona out to several solar radii.
Resolving the \skamid{} gyroresonance layers recovers the coronal field to a few-percent
precision, unlike optical/EUV magnetography, which extrapolates the photospheric field. At
\skalow{} frequencies the Stokes-$V$ diagnostic is less direct, as it encodes both the
emission and propagation regions; absolute field reconstruction there is model-dependent and
requires propagation-aware forward modelling (Section~\ref{sec:crosscutting}).

\subsection{Solar Radio Bursts: Metric to Kilometric}
\label{sec:bursts}

Type~II, III, and IV radio bursts are the principal radio signatures of CMEs,
flare-accelerated electron beams, and trapped energetic-particle populations,
respectively. Their frequency drift rates, fine structures, and polarisation encode the
kinematics and energetics of the driving phenomena as well as the density and
magnetic-field profiles of the medium through which they propagate.
\citet{Kumari01.2026.SKA} review SKAO's capability for comprehensive imaging spectroscopy of
these bursts from metric to decametric wavelengths. \skalow{} will image type~II bursts
driven by CME-associated shocks with sufficient angular resolution to locate the emission
relative to the shock front, while type~III burst imaging will trace electron-beam
trajectories through the corona, constraining the three-dimensional magnetic topology.

\subsection{Probing the Solar Corona with SKA-Mid}
\label{sec:skamid}

At GHz frequencies, the solar corona presents a wealth of fine-scale thermal and
nonthermal structures---active-region loops, prominence--corona interfaces, microflare
sites---that current instruments sample with limited spatial resolution and $uv$-coverage.
\citet{Cheung01.2026.SKA} use forward-modelled MHD simulations to predict the radio appearance
of the corona at \skamid frequencies and demonstrate that AA4 baselines will resolve
structures at scales below 1\,arcsec at 10\,GHz. These capabilities are highly
complementary to forthcoming EUV and optical facilities: the Multi-Slit Solar Explorer
\citep[MUSE;][]{DePontieu2022}, the EUV High-Throughput Spectroscopic Telescope (EUVST),
and the Daniel K.\ Inouye Solar Telescope \citep[DKIST;][]{Rimmele2020}. Joint analysis of
radio, EUV, and photospheric magnetogram data will enable self-consistent modelling of the
thermal, nonthermal, and magnetic properties of the coronal plasma.

\subsection{CME Magnetic Fields}
\label{sec:cmebfield}

The geoeffectiveness of a CME is determined largely by its internal magnetic-field
configuration, a quantity that is extremely difficult to measure remotely.
\citet{Kansabanik01.2026.SKA} describe two complementary SKAO approaches. First,
gyrosynchrotron emission from mildly relativistic electrons within the CME body provides
a spectropolarimetric diagnostic of the magnetic-field strength and orientation out to
$\sim$10\,$\rsun$ \citep{Kansabanik2022}. Second, Faraday rotation of linearly polarised
background sources observed through the CME plasma constrains the line-of-sight magnetic
field at larger heliocentric distances \citep{Kooi2022}. SKAO's wide field of view, large
number of simultaneously observable polarised background sources, and spectropolarimetric
sensitivity will transform CME magnetic-field measurement from isolated case studies into a
systematic, operationally relevant capability.
Because SKAO increases the number of polarised background sightlines per CME from one or
a few to tens or hundreds, Faraday-rotation measurement becomes a tomographic reconstruction
of the internal field---and hence the $B_z$ governing geoeffectiveness---rather than a single
detection.

\begin{table}[t]
\centering
\caption{Comparison of imaging capabilities of selected solar instruments and radio
facilities relevant to SHI science with SKAO. Values are representative; actual
performance depends on observing mode, frequency, and target.
Rows are grouped into dedicated solar facilities (which define the current operational
baseline for solar radio imaging), general-purpose precursor/pathfinder arrays used for
solar work, and SKAO~(AA4). This grouping allows the reader to distinguish what is already
routinely achievable at dedicated solar arrays from the capabilities SKAO uniquely adds.}
\label{tab:instruments}
\begin{tabular}{lcccc}
\hline
\textbf{Instrument} & \textbf{Waveband} & \textbf{Angular Res.} & \textbf{Cadence} & \textbf{Polarimetry}\\
\hline
\multicolumn{5}{l}{\textit{Optical/EUV solar context}}\\
SDO/AIA        & EUV (94--335\,{\AA})   & 1.5\,$''$             & 12\,s            & No \\
SDO/HMI        & 6173\,{\AA}            & 1.0\,$''$             & 45\,s            & Full Stokes \\
DKIST/DL-NIRSP & 500--1500\,nm          & 0.1\,$''$             & $\sim$10\,s      & Full Stokes \\
\hline
\multicolumn{5}{l}{\textit{Dedicated solar radio facilities (current operational baseline)}}\\
EOVSA          & 1--18\,GHz             & $\sim$3\,$''$         & 1\,s             & Dual circular \\
Nan\c{c}ay Radioheliograph & 150--450\,MHz & $\sim$1--4\,$'$ & 0.1--1\,s & Stokes $I$, $V$ \\
MUSER     & 0.4--15\,GHz    & $\sim$1.5--50\,$''$ & $\sim$25\,ms & Dual circular \\
e-Callisto$^c$ & 45--870\,MHz & --- (no imaging) & 0.25\,s  & No \\
\hline
\multicolumn{5}{l}{\textit{General-purpose precursor/pathfinder arrays}}\\
MWA            & 80--300\,MHz           & $\sim$2\,$'$          & 0.5\,s           & Full Stokes \\
LOFAR          & 10--240\,MHz           & $\sim$3\,$''{}^a$     & 0.01\,s          & Full Stokes \\
\hline
\multicolumn{5}{l}{\textit{SKAO}}\\
\textbf{\skalow{}(AA4)} & \textbf{50--350\,MHz}    & \textbf{$\sim$7\,$''$}        & \textbf{$\leq$0.05\,s} & \textbf{Full Stokes} \\
\textbf{\skamid (AA4)} & \textbf{0.35--15.4\,GHz} & \textbf{$\sim$0.04\,$''{}^b$} & \textbf{$\leq$0.05\,s} & \textbf{Full Stokes} \\
\hline
\multicolumn{5}{l}{\footnotesize $^a$International baselines; core-only resolution $\sim$2\,$'$.}\\
\multicolumn{5}{l}{\footnotesize $^b$At 15\,GHz with maximum AA4 baselines.}\\
\multicolumn{5}{l}{\footnotesize $^c$Spectrometer network: high spectral/temporal coverage but non-imaging; listed for context.}
\end{tabular}
\end{table}

Dedicated solar facilities already deliver either high angular resolution over a
restricted band (EOVSA) or fast full-Sun spectral monitoring (Nan\c{c}ay, MUSER, e-Callisto),
and thus define what is \emph{already} achievable. SKAO is unique in combining all of these
capabilities---sub-arcsecond-to-arcsecond imaging, continuous 50\,MHz--15.4\,GHz coverage,
full-Stokes spectropolarimetry, and an order-of-magnitude sensitivity gain---simultaneously.

\section{Heliosphere, Solar Wind, and Energetic Particles}
\label{sec:heliosphere}

Continuing the Sun-to-Earth chain, the eruptions, shocks, and beams of
Section~\ref{sec:solar} now propagate through the turbulent heliosphere, which acts both as
the next physical link and as a propagation filter on all radio waves crossing it.

\subsection{Heliospheric Turbulence and Radio-Wave Propagation}
\label{sec:turbulence}

The turbulent heliospheric plasma imprints observable signatures on radio waves
propagating through it---angular broadening, intensity scintillation, temporal pulse
broadening, and spectral modulation. These propagation effects are both a diagnostic of
the turbulence itself and a source of distortion that must be understood to interpret
solar radio observations correctly. \citet{Chrysaphi01.2026.SKA} present SKAO as a
uniquely powerful probe of multi-scale heliospheric turbulence. The sensitivity gain over
current instruments will allow the intrinsic properties of radio-burst sources to be
cleanly separated from scattering-induced modifications for the first time, resolving
long-standing discrepancies between observed and predicted source sizes. Simultaneously,
the frequency dependence of the scattering signature will constrain the power spectrum
and anisotropy of density fluctuations from spatial scales of tens of kilometres to solar
radii, providing ground truth for heliospheric turbulence models.

\subsection{Interplanetary Scintillation}
\label{sec:ips}

Interplanetary scintillation (IPS)---the rapid intensity fluctuation of compact radio
sources caused by the turbulent solar wind---has been used for decades to map the
three-dimensional structure of the heliosphere. \citet{Morgan01.2026.SKA} discuss, among
other topics, the potential for IPS studies with \skalow{} and \skamid (the same
contribution also treats the ionosphere, Section~\ref{sec:ionosphere}). The dramatically
larger number of compact background sources detectable by SKAO ($>$10$^4$ within a few
degrees of the Sun) will enable tomographic reconstruction of the solar-wind velocity,
density, and turbulence with spatial resolution and cadence far exceeding current IPS
arrays. CME-driven disturbances will be traceable through the inner heliosphere in
near-real time, contributing directly to space-weather forecasting.

\subsection{Angular Broadening and Radio Occultation}
\label{sec:broadening}

The angular broadening of compact radio sources observed at small elongation from the Sun
is a powerful probe of coronal and heliospheric density turbulence.
\citet{PeijinZhang01.2026.SKA} demonstrate that the anisotropy of the angular broadening
kernel encodes information about the magnetic-field topology: field-aligned density
striations produce elongated scatter-broadened images whose orientation traces the
projected field direction. \citet{Raja01.2026.SKA} complement this with a
broader treatment of radio-wave propagation as a diagnostic of the coronal and solar-wind
plasma, including group-delay and Doppler measurements from spacecraft-based and
ground-based occultation experiments. SKAO will dramatically increase the density of
background sources available for occultation and angular-broadening measurements,
enabling continuous monitoring of coronal conditions across the solar cycle.

\subsection{Solar Energetic Particles and Radio Associations}
\label{sec:seps}

Solar energetic particle (SEP) events are a major space-weather hazard, yet the
mechanisms by which particles are accelerated and released from the corona remain poorly
understood. \citet{Morosan01.2026.SKA} address the intimate connection between SEP events and
solar radio bursts: type~III bursts mark electron-beam injection onto open field lines,
type~II bursts trace CME-driven shocks that accelerate particles, and the timing and
spatial relationships between the two provide clues to the acceleration site and release
conditions. SKAO's imaging capability will locate burst sources with arcsecond precision
in the context of CME and shock structures observed by coronagraphs, while the spectral
coverage will track emission from the low corona to the outer heliosphere. When combined
with in-situ particle measurements from Solar Orbiter \citep{Muller2020}, Parker Solar
Probe \citep{Fox2016}, and Aditya-L1, SKAO will close the gap between remote-sensing and
in-situ observations of particle acceleration.

\section{Ionosphere and Geospace Environment}
\label{sec:ionosphere}

The Sun-to-Earth chain closes at the ionosphere. Earth's
ionosphere is both a science target and an instrumental challenge for
low-frequency radio astronomy. \citet{Datta01.2026.SKA} show that radio interferometric
arrays such as \skalow{} measure total electron content (TEC) gradients with sensitivity
exceeding that of Global Navigation Satellite System (GNSS) receivers by an order of
magnitude, enabling characterisation of ionospheric structure on spatial scales from
kilometres to hundreds of kilometres. The mid-latitude location of \skalow{} in Western
Australia (geomagnetic latitude $\sim$$-42^\circ$) provides a complementary perspective
to equatorial and high-latitude GNSS networks, and the continuous all-sky coverage during
astronomical observations yields ionospheric data as a by-product.

\citet{Morgan01.2026.SKA} provide a comprehensive review of methods for observing and
characterising the ionosphere with \skalow, including direction-dependent calibration
solutions that yield spatial maps of differential TEC, forward-modelling using physical
ionospheric models, and the extraction of scintillation statistics. They emphasise the
dual role of ionospheric science: as a research objective in its own right---studying
travelling ionospheric disturbances, medium-scale structures, and the response to
geomagnetic storms---and as a calibration requirement whose solution directly benefits all
SHI (and indeed all low-frequency) science. Accurate ionospheric modelling will be
critical during solar maximum, when ionospheric conditions at the \skalow{} site may be
significantly disturbed, and the Sun itself is the most scientifically productive target.

\section{Stellar Connections and Broader Context}
\label{sec:stellar}

The Sun is the only star whose corona and activity cycle can be spatially resolved, but it
is also just one exemplar of the diverse phenomena displayed by magnetically active cool
stars. \citet{Mohan01.2026.SKA} explore the bridge between solar and stellar radio
astrophysics, describing how SKAO will detect quiescent and flaring radio emission from
M~dwarfs and other active cool stars to distances of tens of parsecs. The direct
SHI relevance is the Sun-as-a-star calibration: the spatially resolved solar measurements
provide the ground truth needed to interpret disk-integrated stellar analogues of the same
processes studied above. SKAO
observations of the Sun as a spatially resolved ``benchmark star'' will calibrate the
interpretation of unresolved stellar detections, while the stellar perspective will place
solar activity in a broader evolutionary and environmental context. As broader
context, the radio environment of a host star---driven by its magnetic
activity, flaring rate, and wind properties---determines the space-weather conditions
experienced by orbiting planets and may influence atmospheric retention and habitability.

\section{Observational, Calibration, and Data-Analysis Frameworks}
\label{sec:calibration}

Realising the SHI science programme with SKAO demands an observational framework that is
in many respects unique among SKA science cases. The Sun is an extremely bright, spatially
extended, rapidly varying source that produces emission spanning many orders of magnitude
in flux density. \citet{Oberoi01.2026.SKA} describe the state-of-the-art in solar radio
observation, calibration, and imaging, drawing on the substantial body of experience
accumulated with SKA precursors and pathfinders---principally the MWA and LOFAR, and more recently also MeerKAT. Key
technical challenges include achieving snapshot dynamic range in excess of $10^5$ against
the bright solar disk, performing direction-dependent calibration in the presence of a
dominant, time-variable extended source, and managing data rates that can exceed
10\,TB\,hr$^{-1}$ in spectropolarimetric imaging modes.

The chapter describes solutions that are being developed for AA4 operations: automated
triggering of burst-mode observations using real-time observations, 
hierarchical data products that balance archival volume against scientific completeness,
and integration with the SKAO Regional Centres (SRCs) for computationally intensive
reprocessing. These frameworks are not SHI-specific in all respects; the calibration and
imaging techniques for wide-field, high-dynamic-range, direction-dependent problems are
broadly applicable across SKA science.

The standard SKAO/SRCNet model, oriented toward archival Observatory Data Products,
provides the necessary calibration, imaging, and archival infrastructure but does not in its
baseline form address solar-specific flux/polarisation calibration, high-cadence
($\lesssim$50\,ms) burst imaging spectroscopy, or the near-real-time products required for
space weather. A dedicated, low-latency SHI processing layer \emph{upstream} of SRCNet is
therefore likely needed; defining its interface is a pre-operations task
(Section~\ref{sec:crosscutting}).

\section{Cross-cutting Themes, Gaps, and the AA4 Outlook}
\label{sec:crosscutting}

Several themes recur across the individual chapters but are not fully addressed by any
single contribution. We highlight these here both to acknowledge their importance and to
motivate future work.

\paragraph{Real-time space-weather operations.}
Many of the science cases described above---CME magnetic-field measurement, type~II burst
tracking, IPS tomography---have direct relevance for operational space-weather
forecasting. The transition from scientific analysis to real-time operational products
requires dedicated pipelines with stringent latency constraints, robust automated quality
control, and interfaces to existing space-weather services. While the potential is
recognised in several chapters, the detailed architecture of a space-weather data stream
from SKAO remains to be developed and would rely on the dedicated SHI processing layer
of Section~\ref{sec:calibration}.

\paragraph{Multi-messenger and multi-mission coordination.}
SKAO will operate contemporaneously with an unprecedented fleet of heliophysics missions:
Solar Orbiter, Parker Solar Probe, Aditya-L1 \citep{Srivastava2021}, PUNCH
\citep{Vourlidas2020}, and PROBA-3, as well as ground-based facilities such as DKIST and
the forthcoming MUSE and EUVST spectrographs. Coordinated observing campaigns that
combine SKAO radio data with remote-sensing and in-situ measurements from these platforms
will maximise the scientific return, but require planning frameworks, data-sharing
protocols, and cross-calibration efforts that are still at an early stage.

\paragraph{Long-term synoptic monitoring.}
The solar cycle is a multi-decadal phenomenon, and many SHI science objectives---the
statistical properties of coronal heating events, the evolution of the heliospheric
turbulence spectrum, ionospheric climatology---benefit from continuous monitoring over a
full activity cycle. Establishing a synoptic solar observing programme within the SKAO
operational model, analogous to the routine monitoring performed by instruments such as
SDO/AIA and GOES, will be important for long-term science return.

\paragraph{AA4 capabilities and the staged roadmap.}
The 16 chapters collectively demonstrate that the AA4 design baseline is sufficient to
achieve transformative SHI science across the full range of topics considered here. The
dense $uv$-coverage at AA4---512 stations for \skalow{} and 197 dishes for \skamid---is
critical for high-fidelity snapshot imaging of the extended, rapidly varying solar source.
Full-Stokes imaging from 50\,MHz to 15.4\,GHz with millisecond time resolution and
$\sim$10\,kHz spectral resolution provides the spectropolarimetric parameter space needed
for the diagnostic techniques described above. Early array assemblies (AA2, AA$^*$) will
already enable significant SHI science---particularly IPS, ionospheric studies, and
observations of the brightest solar bursts---and will serve as essential pathfinding
stages for commissioning solar observing modes and validating calibration strategies.

\paragraph{Community infrastructure.}
Delivering the SHI science programme at scale will require community-developed analysis
pipelines, simulation tools for forward modelling of radio observables from MHD models,
and integration with the SRC computing infrastructure. Investment in SHI-specific software
tools---building on existing efforts for MWA and LOFAR solar data---and in training the
next generation of solar radio astronomers are essential enabling activities.

\paragraph{Consistent forward modelling across SKA-Low and SKA-Mid.}
Forward modelling of radio observables is more developed for \skamid{}
\citep{Cheung01.2026.SKA} than for \skalow, where scattering, angular broadening, and
low-frequency magnetography rely on propagation effects not yet embedded in a common,
polarisation-aware framework. Developing a \skalow{} capability consistent with the
\skamid{} one---and stating the associated assumptions and limitations---is a priority for
the SHI programme.

\section{Summary}
\label{sec:summary}

The 16 topical chapters presented in this volume demonstrate that SKAO will be a
transformative facility for solar, heliospheric, and ionospheric physics. From resolving
the nonthermal signatures of nanoflare-scale coronal heating events
\citep{Mondal01.2026.SKA} to mapping the three-dimensional magnetic-field structure of CMEs
\citep{Kansabanik01.2026.SKA}, from probing multi-scale heliospheric turbulence
\citep{Chrysaphi01.2026.SKA} to characterising ionospheric disturbances with unprecedented
spatial resolution \citep{Datta01.2026.SKA, Morgan01.2026.SKA}, the science cases
collectively define a programme of enormous breadth and depth tracing a single
end-to-end Sun-to-Earth system.

The design baseline AA4 capabilities---sub-arcsecond resolution, full-Stokes
spectropolarimetric imaging, sensitivity gains of an order of magnitude over existing
facilities, and continuous spectral coverage from 50\,MHz to 15.4\,GHz---are well matched
to the requirements articulated in each chapter. Cross-cutting challenges remain, notably
in real-time space-weather data delivery, multi-mission coordination, long-term synoptic
monitoring, and community software infrastructure; these represent important areas for
further development as SKAO moves toward early science operations and the full realisation
of AA4.

The SHI Science Working Group looks forward to a new era in solar and heliospheric radio
physics with the SKAO, one that will deepen our understanding of fundamental plasma
processes and strengthen our ability to predict and mitigate the space-weather hazards
that increasingly affect our technological society.

\bibliographystyle{abbrvnat-maxbibnames4}
\bibliography{SHI_overview}

\end{document}